\documentclass{aa}

\usepackage{txfonts}
\usepackage{graphicx}
\usepackage{natbib}
\bibpunct{(}{)}{;}{a}{}{,}

\begin{document}

\title{YSO accretion shocks: magnetic, chromospheric or stochastic flow effects
  can suppress fluctuations of X-ray emission}

\author{
  T.~Matsakos\inst{1,2,3}, J.-P.~Chi\`eze\inst{2}, C.~Stehl\'e\inst{3},
  M.~Gonz\'alez\inst{4}, L.~Ibgui\inst{3}, L.~de~S\'a\inst{2,3},
  T.~Lanz\inst{5}, S.~Orlando\inst{6}, R.~Bonito\inst{7,6},
  C.~Argiroffi\inst{7,6}, F.~Reale\inst{7,6} \and G.~Peres\inst{7,6}
}

\authorrunning{T. Matsakos et al.}
\titlerunning{Suppressing the periodic emission of YSO accretion shocks}

\institute{
  CEA, IRAMIS, Service Photons, Atomes et Mol\'ecules, 91191 Gif-sur-Yvette,
    France \and
  Laboratoire AIM, CEA/DSM - CNRS - Universit\'e Paris Diderot, IRFU/Service
    d'Astrophysique, CEA Saclay, Orme des Merisiers, 91191 Gif-sur-Yvette,
    France \and
  LERMA, Observatoire de Paris, Universit\'e Pierre et Marie Curie and CNRS, 5
    Place J. Janssen, 92195 Meudon, France \and
  Universit\'e Paris Diderot, Sorbonne Paris Cit\'e, AIM, UMR 7158, CEA, CNRS,
    91191 Gif-sur-Yvette, France \and
  Laboratoire Lagrange, Universit\'e de Nice-Sophia Antipolis, CNRS,
    Observatoire de la C\^ote d'Azur, 06304 Nice cedex 4, France \and
  INAF - Osservatorio Astronomico di Palermo, Piazza del Parlamento 1, 90134
    Palermo, Italy \and
  Dipartimento di Fisica e Chimica, Universit\`a degli Studi di Palermo, Piazza
    del Parlamento 1, 90134 Palermo, Italy
}

\date{Received ?? / Accepted ??}

\abstract{
  Theoretical arguments and numerical simulations of radiative shocks produced
  by the impact of the accreting gas onto young stars predict quasi-periodic
  oscillations in the emitted radiation.
  However, observational data do not show evidence of such periodicity.
}{
  We investigate whether physically plausible perturbations in the accretion
  column or in the chromosphere could disrupt the shock structure influencing
  the observability of the oscillatory behavior.
}{
  We performed local 2D magneto-hydrodynamical simulations of an accretion shock
  impacting a chromosphere, taking optically thin radiation losses and thermal
  conduction into account.
  We investigated the effects of several perturbation types, such as clumps in
  the accretion stream or chromospheric fluctuations, and also explored a wide
  range of plasma-$\beta$ values.
}{
  In the case of a weak magnetic field, the post-shock region shows chaotic
  motion and mixing, smoothing out the perturbations and retaining a global
  periodic signature.
  On the other hand, a strong magnetic field confines the plasma in flux tubes,
  which leads to the formation of fibrils that oscillate independently.
  Realistic values for the amplitude, length, and time scales of the
  perturbation are capable of bringing the fibril oscillations out of phase,
  suppressing the periodicity of the emission.
}{
  The strength of a locally uniform magnetic field in YSO accretion shocks
  determines the structure of the post-shock region, namely, whether it will be
  somewhat homogeneous or if it will split up to form a collection of fibrils.
  In the second case, the size and shape of the fibrils is found to depend
  strongly on the plasma-$\beta$ value but not on the perturbation type.
  Therefore, the actual value of the protostellar magnetic field is expected to
  play a critical role in the time dependence of the observable emission.
}


\maketitle

\section{Introduction}

The accretion process of young stellar objects (YSOs) is believed to take place
by funneled streams that originate in the surrounding disk and flow along the
field lines of the protostellar magnetosphere
\citep{Koe91,Har94,Rom03,Rom04,Rom08}.
The gas falls onto the surface of the central body with a supersonic velocity,
close to the free-fall speed.
Its impact on the chromosphere produces strong shocks of a few million Kelvin,
a phenomenon that is observable in X-rays
\citep{Ulr76,Gul94,Cal98,Lam98,Lam99,Bo07a}.
Detailed observations of young stars have detected such emission
\citep{Kas02,Sch05,Gue06,Arg07,Rob07,Bri10,Dup12}, which has to be carefully
distinguished from the coronal one.
The presence of an accretion shock is probed by a denser
($n_\mathrm{e} \ga 10^{11}\,\mathrm{cm}^{-3}$) and cooler
($T \sim 2$--$5\mathrm{MK}$) plasma, as compared to the more tenuous, at least
by one order of magnitude, and hotter corona.

\begin{figure}
  \resizebox{\hsize}{!}{\includegraphics{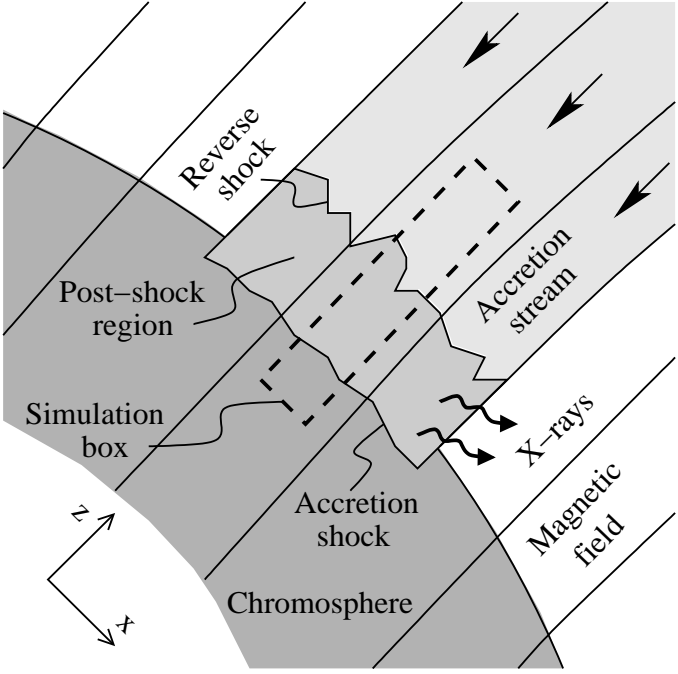}}
  \caption{A schematic low plasma-$\beta$ configuration of an accretion shock.
    A strong magnetic field penetrates the chromosphere and confines the plasma
    in flux tubes.
    The accretion stream impacts on the surface and pushes the chromosphere down
    to the point where the ram pressure is equal to the stellar pressure.
    The computational box for the simulations presented in this paper is
    indicated with a dashed line.}
  \label{fig:sketch}
\end{figure}

The time dependence of the accretion shock's structure is determined by
radiative processes, as well as by the topology and strength of the magnetic
field \citep{Sac08,Sac10,Kol08,Orl10}.
The typical field of protostars can be strong enough, on the order of
$\mathrm{kG}$ \citep{Joh07,Don07}, such that it confines the infalling plasma
inside magnetic flux tubes and prevents it from escaping sideways.
Upon impact, a shock is transmitted into the chromosphere, and a reverse shock
propagates along the stream.
The simplest local configuration is to assume that the chromosphere is
horizontal and threaded by a perpendicular uniform field, as depicted in
Fig.~\ref{fig:sketch}.
Material accumulates vertically and the thickness of the shocked region
increases, building up a hot slab of a few million Kelvin.
Considering the approximation of an optically thin stream, the hot slab radiates
its energy away and collapses within the characteristic cooling time
($\sim$$10$--$10^3\,\mathrm{s}$).
This process is expected to repeat quasi-periodically, giving rise to a
pulsating emission.
Such a cycle has been demonstrated and examined extensively in 1D numerical
studies of YSO accretion shocks \citep{Sac08,Kol08,Sac10}.
In fact, it is a general feature of radiative shocks that the dependence of the
cooling function on the temperature may trigger cooling instabilities.
In turn, this may lead to the recurrent formation and collapse of the reverse
shock, both in the hydrodynamical regime
\citep[e.g.][]{Lan81,Che82,Ra05a,Sut03,Mig05} and in the presence of a
transverse magnetic field \citep[e.g.][]{Tot93,Ra05b}.

However, recent observational data of accreting YSOs do not show evidence for
such an oscillatory structure \citep{Dra09,Gue10}.
In particular, \citet{Dra09} analyzed the time series of X-ray observations of
TW Hya and excluded the presence of periodicity over the frequency range
$0.0001$--$6.811\,\mathrm{Hz}$, which they argue to be the most likely window
(the upper limit of the relative amplitude, $A$, is $0.033$ for the interval
$0.0001$--$0.192\,\mathrm{Hz}$ and $A = 0.056$ for
$0.192$--$6.811\,\mathrm{Hz}$).
Moreover, \citet{Gue10} reach a similar conclusion having examined optical and
UV data of TW Hya and AA Tau over the frequency ranges $0.02$--$3\,\mathrm{Hz}$
($A = 0.001$) and $0.02$--$50\,\mathrm{Hz}$ ($A = 0.005$), respectively.
The absence of periodicity can be attributed to several possible reasons.
For instance, by taking radiation transfer effects into account, the cooling of
the post-shock region may change and lead to a stable shock stalled inside the
chromosphere (Chi\`eze et al. in preparation).
On the other hand, complex magnetic field topologies can invalidate the 1D
approximation and thus its periodic behavior (\citeauthor{Orlsu} submitted).
In a similar context, simulations performed in the high plasma-$\beta$ regime
(ratio of the thermal pressure over the magnetic one) indicate a rather smooth
emission, thanks to the shocked material that escapes sideways from the
accretion column and deflates the hot slab \citep{Orl10}.
We also note that \citet{Sut03} explored cooling instabilities of radiative
shocks in the pure hydro-dynamical 2D case and found that the extra dimension
together with small perturbations lead to inhomogeneous shocks.
Their filamentary structure enhanced the cooling efficiency and generated
turbulence that may introduce ambiguities in the derivation of velocity
profiles.

On the other hand, several authors \citep[e.g.][]{Gul96,Saf98,Bo07b} have
pointed out that accretion streams of young stars are clumped and inhomogeneous,
while until now only homogeneous accretion flows have been modeled.
A detailed multi-dimensional magneto-hydrodynamical (MHD) modeling of a density
structured accretion flow impacting on the stellar surface is therefore required
to investigate the complex structure and stability of the hot slab for an
inhomogeneous stream.

The 1D accretion model that is often found in the literature is based on the
assumption of a locally strong and uniform field.
By considering typical accretion stream velocities and densities, namely
$v_\mathrm{acc} = 100$--$500\,\mathrm{km\,s^{-1}}$ and
$\rho_\mathrm{acc} = 10^{-13}$--$10^{-11}\,\mathrm{g\,cm}^{-3}$
\citep[][and references therein]{Sac10}, and a local value of the magnetic field
of $B = 0.1$--$2\,\mathrm{kG}$ in the impact region, the plasma-$\beta$ value in
the post-shock region\footnote{In the following, the plasma-$\beta$ value will
always refer to the post-shock region.} is on the order
\begin{equation}
  \beta_\mathrm{sh} \equiv \frac{P_\mathrm{sh}}{B^2/8\pi}
    \simeq \frac{\rho_\mathrm{acc} v_\mathrm{acc}^2}{B^2/8\pi}
    \sim 10^{-2}\!\! - \!\!10^{2}\,,
\end{equation}
where we have approximated the thermal pressure of the shocked region,
$P_\mathrm{sh}$, with the ram pressure of the infalling gas.
For the parameter space corresponding to $\beta \lesssim 1$, the plasma is
constrained within vertical flux tubes in which case the 1D approximation is
justified.

In this paper, we extend the above configuration and we perform 2D MHD
simulations.
During the evolution, we introduce either density perturbations in the accretion
stream, as it is the case for a clumpy flow, or pressure variabilities in the
chromosphere, similar to what is observed in the sun.
We study the time dependence of the shock structure for several plasma-$\beta$
values, looking for multi-dimensional mechanisms that could disrupt or even
suppress the oscillation.
We note that the presence of a strong magnetic field constrains significantly
the plasma dynamics; for instance it can prevent the generation of large scale
chaotic motion such as that observed in \citet{Sut03}.

The paper is structured as follows.
Section~\ref{sec:setup} decribes the numerical setup and lists the simulation
models.
Section~\ref{sec:analytical} derives simple analytical expressions to estimate
the effects that perturbations can have on the temporal shock evolution.
Section~\ref{sec:results} presents our numerical results and discusses their
relevance to real accretion shocks.
Finally, Sect.~\ref{sec:conclusions} summarizes our work and reports our
conclusions.

\section{Theoretical framework}
  \label{sec:setup}

\subsection{The MHD equations}

The MHD perfect gas equations, including the appropriate terms suited for the
study of accretion shocks, are
\begin{equation}
  \frac{\partial\rho}{\partial t} + \nabla \cdot (\rho \vec v) = 0\,,
\end{equation}
\begin{equation}
  \frac{\partial\vec v}{\partial t} + (\vec v \cdot \nabla)\vec v
    + \frac{1}{4\pi\rho}\vec B \times (\nabla \times \vec B)
    + \frac{1}{\rho}\nabla P = \vec g\,,
  \label{eq:v_eq}
\end{equation}
\begin{equation}
  \frac{\partial P}{\partial t} + \vec v \cdot \nabla P
    + \gamma P \nabla \cdot \vec v = -(\gamma - 1)\big(\nabla\cdot\vec F
    + n_\mathrm{e}n_\mathrm{H}\Lambda\big)\,,
  \label{eq:pr_eq}
\end{equation}
\begin{equation}
  \frac{\partial\vec B}{\partial t} + \nabla \times (\vec B \times \vec v)
    = 0\,.
\end{equation}
The equations are written in cgs units for the variables, $\rho$, $P$, and
$\vec v$, which denote the gas density, pressure, and velocity, respectively.
The magnetic field, $\vec B$, also satisfies the condition
$\nabla\cdot\vec B = 0$.
The ratio of specific heats is $\gamma = 5/3$ and $\vec g$ represents gravity.
The thermal conduction flux is given by
\begin{equation}
  \vec F = \frac{F_\mathrm{sat}}{F_\mathrm{sat} + |\vec F_\mathrm{Spi}|}
    \vec F_\mathrm{Spi}\,,
\end{equation}
where $F_\mathrm{sat} = 5\rho c_s^3$ is the saturated flux \citep{Cow77} and
$c_s$ the sound speed.
The term $\vec F_\mathrm{Spi}$ consists of two components, one along ($\|$) and
one across ($\bot$) the magnetic field lines \citep{Spi62}, given by
\begin{equation}
  \vec F_\mathrm{Spi} = \kappa_\|\frac{(\vec B\cdot\nabla T)\vec B}{\vec B^2}
    + \kappa_\bot\left(\nabla T
    - \frac{(\vec B\cdot\nabla T)\vec B}{\vec B^2}\right)\,,
\end{equation}
where $T$ is the temperature, $n_\mathrm{H}$ is the number density of hydrogen
atoms, and $k_\mathrm{B}$ is the Boltzmann constant.
The thermal conduction coefficients are given by
$\kappa_\| = -9.2\times10^{-7}T^{5/3}$ and
$\kappa_\bot = -3.3\times10^{-16}n_\mathrm{H}^2/(T^{1/2}\vec B^2)$,
respectively, both of which are in units of
$\mathrm{erg\,s}^{-1}\,\mathrm{K}^{-1}\,\mathrm{cm}^{-1}$.
A mean atomic weight of $\mu = 1.2$ is considered, i.e.
$\rho = \mu n_\mathrm{H}m_\mathrm{H}$, where $m_\mathrm{H}$ is the hydrogen
atomic mass.
The electron number density is denoted with $n_\mathrm{e}$ and the gas is
assumed to be always fully ionized.
The latter is a valid approximation for the post-shock region, the cooling of
which plays a key role in this study.
The pre-shocked material is cold enough such that the energy losses are
negligible and the actual value of $n_\mathrm{e}$ is not important there.

\begin{figure}
  \input{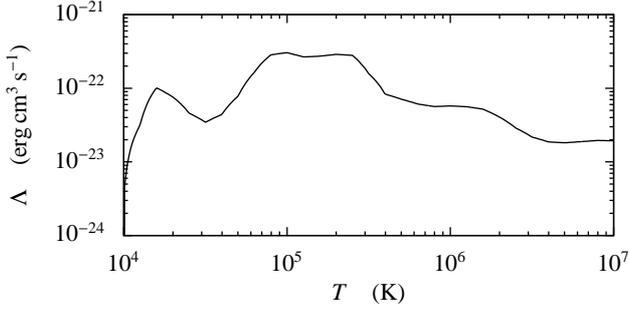}
  \caption{The adopted cooling function, $\Lambda$, as a function of the
    temperature.
    The negative slope in the temperature range $T = 10^5$--$10^7\,\mathrm{K}$
    is triggering cooling instabilities: the more the plasma cools the higher
    the energy losses are.}
  \label{fig:coolfunc}
\end{figure}
The cooling function, $\Lambda$, is adopted from \citet{Orl10} and is shown in
Fig.~\ref{fig:coolfunc}.
It depends on the temperature and represents radiative losses of an optically
thin plasma.
The profile was derived with the PINTofALE spectral code \citep{Kas00} using the
APED V1.3 atomic line database \citep{Smi01} and considering metal abundances
$0.5$ times the solar ones \citep{Tel07}.
This abundance has been choosen because some observations tend to indicate that
the accretion flow is depleted in metals \citep{Ste04,Dra05}.
Finally, in order to prevent the chromosphere from cooling, we define a
temperature threshold, $T_\mathrm{cut} = 10^4\,\mathrm{K}$, below which
$\Lambda$ is set to zero.

Finally, we make use of a flow tracer, $Q$, which is a passive scalar that obeys
the advection equation: $\partial Q/\partial t + \vec v\cdot\nabla Q = 0$.
It does not interfere with the dynamics but simply follows the motion of the
plasma and helps visualize its path.

\subsection{Numerical setup}

The simulations are carried out with PLUTO\footnote{Freely available at
\texttt{http://plutocode.ph.unito.it}}, a numerical code for computational
astrophysics \citep{Mig07,Mig12}.
We adopt the ROE solver for the integration and we apply second order accuracy
on both space and time.

The configuration is simulated in 2D cartesian coordinates\footnote{The unit
vectors are denoted with $\hat x$ and $\hat z$.}, $(x,\,z)$, neglecting surface
curvature and stellar rotation, as shown in Fig.~\ref{fig:sketch}.
We focus on the interior of the stream in order to remain as close as possible
to the 1D problem.
Therefore, we avoid influence from the effects at the interface between the
corona and the funnel flow \citep[e.g.][]{Orl10}.
A chromosphere is initialized on the lower part of the computational box,
underneath a hot corona \citep[e.g.][]{Sac08,Sac10,Orl10}.
Gravity is assumed uniform, $\vec g = -(GM/R^2)\hat z$, with the mass and radius
of the star equal to those of MP Mus, i.e. $M = 1.2M_{\sun}$ and
$R = 1.3R_{\sun}$ \citep{Arg07}.
For simplicity, both the chromosphere and the corona are assumed isothermal,
with temperatures, $T_\mathrm{chr} = 10^4\,\mathrm{K}$ and
$T_\mathrm{cor} = 10^6\,\mathrm{K}$, respectively.
The pressure and density profiles of each one have an exponential dependence
such that the configuration is in equilibrium and pressure is continuous.
We note that during the first steps of the simulations the corona is readily
engulfed by the accretion stream and hence its presence is not important for
this study.
Finally, a uniform vertical magnetic field, $\vec B = B_0\hat z$, is initialized
throughout the box and is evolved self-consistently during the simulation.

The accretion stream is not initially present in the domain but enters from the
top.
In particular, we prescribe the properties of the infalling plasma on the upper
boundary, namely, velocity $v_\mathrm{acc} = 500\,\textrm{km\,s}^{-1}$, number
density $n_\mathrm{acc} = 10^{11}\,\textrm{cm}^{-3}$ and temperature
$T_\mathrm{acc} = 10^3\,\mathrm{K}$ \citep[e.g.][]{Bar08}.
These values, along with $B$, are kept fixed during the simulation, but see also
Sect.~\ref{sec:models} for more details on the applied perturbation.
At the ghost cells of the bottom boundary, we specify zero velocities, we set
the magnetic field equal to its initial uniform value, and we apply the
isothermal equilibrium profiles for the density and pressure.
Moreover, by fixing the physical quantities to a given value implies that MHD
waves might be reflected.
We have investigated the effect of such boundary conditions and found that they
can be neglected.
This can be seen from the reference simulation \texttt{NoPert} which reproduces
the 1D oscillatory behavior reported in the literature (see
Sect.~\ref{sec:1Devolution}).
Lastly, at the left and right edges of the box we apply periodic boundary
conditions.

The computational domain spans $x = [0.0, 0.5\times10^{-3}]\,R_{\sun}$ in the
horizontal direction and $z = [0.0, 7.7\times10^{-2}]\,R_{\sun}$ in the
vertical, and is resolved by a static grid of $[128\times2048]$ cells.
We have not made use of Adaptive Mesh Refinement (AMR) since the nature of the
problem is such that most parts of the computational domain require a high level
of refinement.
Besides, we also avoid the spurious perturbations that could be introduced by a
dynamically evolving grid and would interfere with those applied physically in
the system.
Note that due to the significant elongation of the simulation box, all spatial
distribution plots expand the horizontal direction by a factor of $2$ and hence
the figures look vertically compressed.
The final time of the simulations is on the order of $5$ or $10\,\mathrm{ks}$,
depending on the case under consideration, and corresponds approximately to
$5$--$10$ shock oscillations of the 1D case.
Each simulation requires from a few to several thousands hours of computational
time to complete, due to constraints imposed on the time step by the optically
thin radiation cooling and thermal conduction.
This prevented us from exploring the simulations for longer times.

\subsection{Models and the applied perturbations}
  \label{sec:models}

\begin{figure*}
  \centering 
  \begin{tabular}{ccc}
    \resizebox{0.3\hsize}{!}{\includegraphics{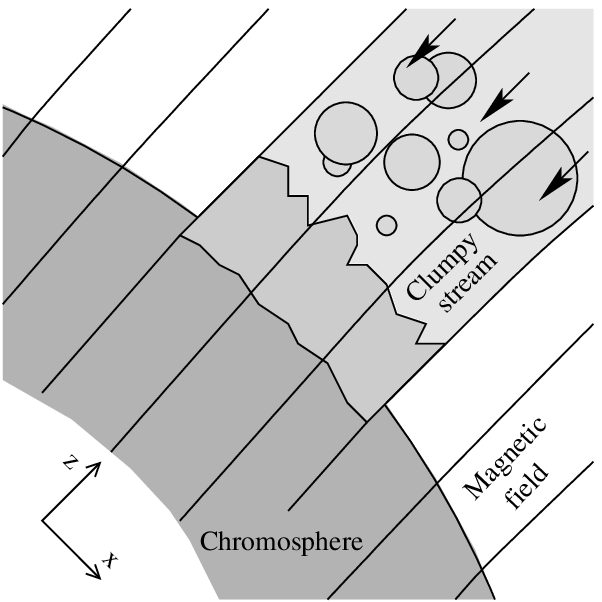}} &
    \resizebox{0.3\hsize}{!}{\includegraphics{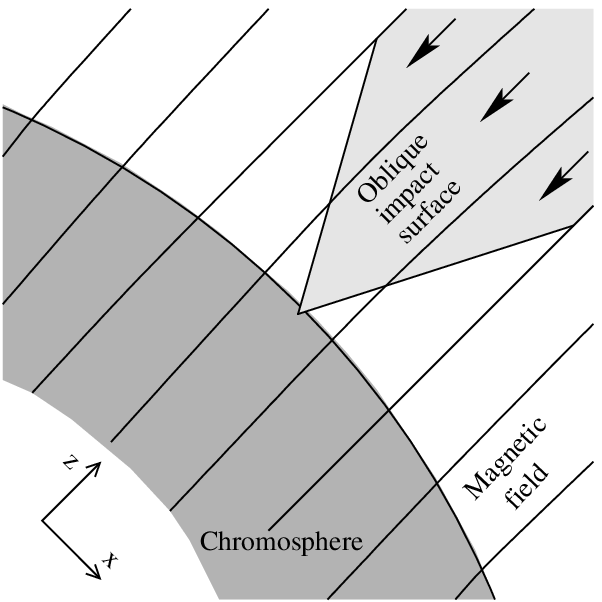}} &
    \resizebox{0.3\hsize}{!}{\includegraphics{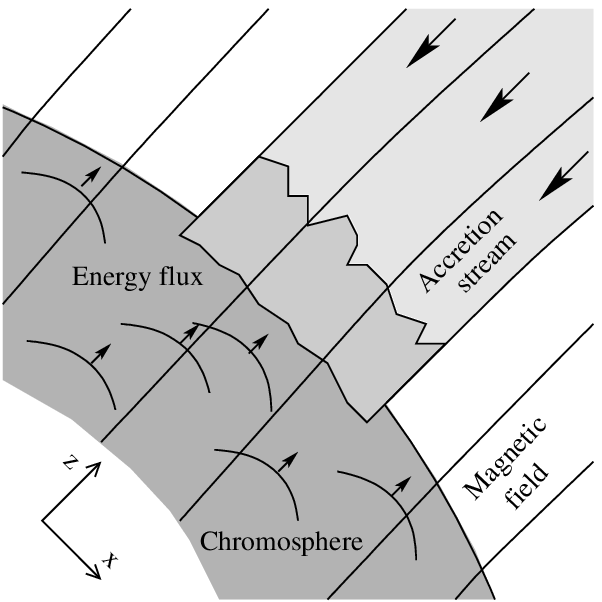}} \\
  \end{tabular}
  \caption{The types of perturbation considered: a clumpy structure of the
    accretion stream (left; models \texttt{Rnd\#},
    \texttt{IsSm\#}/\texttt{IsLg\#}, \texttt{ClSm\#}/\texttt{ClLg\#}), an
    oblique impacting surface (middle; model \texttt{Obl500}), and an energy
    flux present in the chromosphere (right; models \texttt{ChrFlx\#}).}
  \label{fig:perturbation}
\end{figure*}
\begin{table*}
  \caption{The numerical models.
    Their code names summarize the type of perturbation and the strength of the
    magnetic field.}
  \label{tab:models}
  \centering 
  \begin{tabular}{lccccccc}
    \hline
    \hline
    Name & $B_0\,(\mathrm{G})$ & Amplitude & Length scale & Time scale
      & Location & Perturbation description & Final time \\
    \hline
    \texttt{Rnd20}  & $20$  & $0.01$ & $25\,\mathrm{km}$ & $0.05\,\mathrm{s}$
      & Accretion stream & Random density perturbations & $7\,\mathrm{ks}$ \\
    \texttt{Rnd50}  & $50$  & $0.01$ & $25\,\mathrm{km}$ & $0.05\,\mathrm{s}$
      & Accretion stream & Random density perturbations & $7\,\mathrm{ks}$ \\
    \texttt{Rnd100} & $100$ & $0.01$ & $25\,\mathrm{km}$ & $0.05\,\mathrm{s}$
      & Accretion stream & Random density perturbations & $7\,\mathrm{ks}$ \\
    \texttt{Rnd250} & $250$ & $0.01$ & $25\,\mathrm{km}$ & $0.05\,\mathrm{s}$
      & Accretion stream & Random density perturbations & $7\,\mathrm{ks}$ \\
    \texttt{Rnd500} & $500$ & $0.01$ & $25\,\mathrm{km}$ & $0.05\,\mathrm{s}$
      & Accretion stream & Random density perturbations & $7\,\mathrm{ks}$ \\
    \hline
    \texttt{IsSm20}  & $20$  & $10$ & $300\,\mathrm{km}$
      & $30\,\mathrm{s}$  & Accretion stream & Isolated small clumps
      & $4\,\mathrm{ks}$ \\
    \texttt{IsSm500} & $500$ & $10$ & $300\,\mathrm{km}$
      & $30\,\mathrm{s}$  & Accretion stream & Isolated small clumps
      & $4\,\mathrm{ks}$ \\
    \texttt{IsLg20}  & $20$  & $10$ & $1000\,\mathrm{km}$
      & $2\,\mathrm{min}$ & Accretion stream & Isolated large clumps
      & $4\,\mathrm{ks}$ \\
    \texttt{IsLg500} & $500$ & $10$ & $1000\,\mathrm{km}$
      & $2\,\mathrm{min}$ & Accretion stream & Isolated large clumps
      & $4\,\mathrm{ks}$ \\
    \hline
    \texttt{ClSm20}  & $20$  & $10$ & $300\,\mathrm{km}$
      & $0.5\,\mathrm{s}$ & Accretion stream & Stream of small clumps
      & $4\,\mathrm{ks}$ \\
    \texttt{ClSm500} & $500$ & $10$ & $300\,\mathrm{km}$
      & $0.5\,\mathrm{s}$ & Accretion stream & Stream of small clumps
      & $4\,\mathrm{ks}$ \\
    \texttt{ClLg20}  & $20$  & $10$ & $1000\,\mathrm{km}$
      & $1.5\,\mathrm{s}$ & Accretion stream & Stream of large clumps
      & $4\,\mathrm{ks}$ \\
    \texttt{ClLg500} & $500$ & $10$ & $1000\,\mathrm{km}$
      & $1.5\,\mathrm{s}$ & Accretion stream & Stream of large clumps
      & $4\,\mathrm{ks}$ \\
    \hline
    \texttt{Obl500} & $500$ & - & $3500\,\mathrm{km}$ & $15\,\mathrm{min}$
      & Accretion stream & Oblique impacting surface & $7\,\mathrm{ks}$ \\
    \hline
    \texttt{ChrFlx20}  & $20$  & $10F_{\sun}$ & $3500\,\mathrm{km}$
      & $15\,\mathrm{min}$ & Chromosphere & Energy flux & $5\,\mathrm{ks}$ \\
    \texttt{ChrFlx100} & $100$ & $10F_{\sun}$ & $3500\,\mathrm{km}$
      & $15\,\mathrm{min}$ & Chromosphere & Energy flux & $5\,\mathrm{ks}$ \\
    \texttt{ChrFlx500} & $500$ & $10F_{\sun}$ & $3500\,\mathrm{km}$
      & $15\,\mathrm{min}$ & Chromosphere & Energy flux & $5\,\mathrm{ks}$ \\
    \hline
    \texttt{NoPert} & $20$ & - & - & - & - & No perturbation
      & $10\,\mathrm{ks}$ \\
    \hline
  \end{tabular}
\end{table*}
Table~\ref{tab:models} lists the simulations that have been carried out, each
one assuming a different type of perturbation which is visualized in
Fig.~\ref{fig:perturbation}.
In general, we investigate five broad classes of models, a) one that introduces
random density perturbations in the accretion stream, b) one that assumes
isolated clumps in the infalling material, c) one that considers a fully clumped
stream, d) one having an oblique surface of first impact, and e) one that
includes variability in the chromosphere.
For the density fluctuations of the infalling gas, we explore the minimum and
maximum spatial scales that are defined by the resolution and size of the
computational box, respectively.

In particular, the models \texttt{Rnd\#} consider random density perturbations
of $\sim$$1\%$ that take place at the cell level, or equivalently, at a scale of
$\sim$$25\,\mathrm{km}$.
This is done by prescribing the density at the upper boundary in the following
way:
\begin{equation}
  \rho = \rho_\mathrm{acc} + \delta\rho_\mathrm{acc}
    = (1 + q_\mathrm{rand})\rho_\mathrm{acc}\,,
\end{equation}
where $q_\mathrm{rand}$ is a random number that follows a uniform distribution
in the range $[-0.01,\,0.01]$.
The value of $q_\mathrm{rand}$ is different for each ghost cell and it is
randomized repeatedly with a period of
$\Delta t_\mathrm{rand} = \Delta z/v_\mathrm{acc}\simeq0.05\,\mathrm{s}$.
This interval corresponds to the time required for the stream to travel a
distance equivalent to the cell height, $\Delta z$.

The models \texttt{IsSm\#} and \texttt{IsLg\#} consider isolated clumps in the
accretion stream, which are incorporated by adopting a Gaussian profile with a
constant effective width of $\sigma_0$, a randomized location $x_0$, and a
randomized time of appearance $t_0$.
Specifically, the density at the upper boundary condition is set as
\begin{eqnarray}
  \rho &=& \rho_\mathrm{acc} + \delta\rho_\mathrm{acc} = \nonumber \\
    &=& \left[1 + q_\mathrm{cl}
    \exp{\left(-\frac{(x - x_0)^2 + v_\mathrm{acc}^2(t - t_0)^2}
    {\sigma_0^2}\right)}\right]\rho_\mathrm{acc}\,,
\end{eqnarray}
where $q_\mathrm{cl} = 10$ is the peak density contrast and $t_0$ is the time
that the clump enters the computational box.
The value of $t_0$ is chosen in such a way that the clump is initialized far
away from the boundary.
New values are given to $x_0$ and $t_0$ every $\Delta t_\mathrm{cl}$, a time
interval large enough to ensure that the boundary density is equal to
$\rho_\mathrm{acc}$ before each new randomization.
Note that a small randomization applies also to $\Delta t_\mathrm{cl}$ in order
to avoid an exactly periodic appearance of clumps.
For the case of large clumps, \texttt{IsLg\#}, the parameters are
$\sigma_0 = 1000\,\mathrm{km}$ and $\Delta t_\mathrm{cl} = 2\,\mathrm{min}$,
whereas for the case of small clumps, \texttt{IsSm\#}, the parameters are
$\sigma_0 = 300\,\mathrm{km}$ and $\Delta t_\mathrm{cl} = 30\,\mathrm{s}$.

The models \texttt{ClSm\#} and \texttt{ClLg\#} describe a fully clumped stream
using the same gaussian distribution for the overdensities as for the isolated
clumps, with the difference that the clumps appear so frequently that they may
overlap.
In particular, new values are given to $x_0$ and $t_0$ every
$\Delta t_\mathrm{cl} = \sigma_0/v_\mathrm{acc}$, which is equivalent to a
vertical clump spacing comparable to their radius.
Moreover, in order to simulate an accretion stream that is overall described by
the value $\rho_\mathrm{acc}$, we set the peak density of the gaussian
distribution equal to that value.
For numerical reasons, we apply a background uniform density which is $10$ times
lower than $\rho_\mathrm{acc}$.

For the case of an oblique impacting surface, the assumption of periodic
boundaries in the horizontal direction is only valid in the low plasma-$\beta$
regime, where the magnetic field can effectively confine the plasma and prevent
it from escaping sideways.
Therefore, we explore just one model for a large value of $\vec B$,
\texttt{Obl500}.
To achieve an oblique impacting surface, we define the accreting gas variables
only at the part of the top boundary that satisfies the condition:
\begin{equation}
  \left|x - \frac{x_\mathrm{max}}{2}\right|
    < \frac{t}{t_\mathrm{fill}}\frac{x_\mathrm{max}}{2}\,,
\end{equation}
where $x_\mathrm{max}$ is the width of the computational box and
$t_\mathrm{fill} = 15\,\mathrm{min}$ is the time that the stream needs to fill
up the whole domain.
At each time step, the cross section of the stream increases giving a conical
shape to the impacting surface, with the value of $t_\mathrm{fill}$ effectively
controlling its aperture.

On the other hand, models \texttt{ChrFlx\#} consider a homogeneous accretion
stream, and instead perturb the system from the lower edge of the computational
box, i.e. the chromosphere.
In particular, we introduce an energy flux, of a similar type to what is
observed in the sun, by specifying a pressure variation on the bottom boundary
that depends on both time and space:
\begin{equation}
  P = P_\mathrm{chr} + \delta P_\mathrm{chr} = P_\mathrm{chr}
    + \frac{F_\mathrm{chr}}{c_\mathrm{chr}}
    \sin\left(\frac{2\pi t}{t_\mathrm{chr}} - \frac{2\pi x}{L}\right)\,.
\end{equation}
In the above expression, $c_\mathrm{chr}$ is the chromospheric sound speed,
$F_\mathrm{chr} = 5\times10^9\,\mathrm{erg}\,\mathrm{cm}^{-2}\,\mathrm{s}^{-1}$
is the energy flux assumed, $t_\mathrm{chr} = 15\,\mathrm{min}$ is the
periodicity of the variation, and $L = 3500\,\mathrm{km}$ its length scale
\citep[e.g.][]{Boh84,Jud06,Bel09,Bec09,Bec13}.
We have chosen a value for $F_\mathrm{chr}$ that is an order of magnitude
larger than the solar one because young stars are in general expected to be more
active than the sun.
In addition, note that the assumed periodicity time scale of the energy flux is
a few times larger than that of the shock oscillation time, which suggests that
a value closer to the solar one, i.e. a few minutes, would perturb more strongly
the system.
We also apply an appropriate variability to the density at the bottom boundary
such that the chromosphere remains isothermal and hence we avoid thermal
conduction effects that could diffuse the perturbation.
We note that a purely temporal fluctuation, equivalent to a 1D problem, is also
relevant to the study of its effect on the shock oscillations.
However, such cases are explored in de S\'a et al. (in preparation) and hence we
do not discuss them here.

The simulations are performed in both a high and a low plasma-$\beta$ regime,
and in particular for magnetic fields ranging between $20\,\textrm{G}$ and
$500\,\textrm{G}$, or equivalently for post-shock $\beta$ values from
$\beta \la 100$ to $\beta \ga 0.01$.
This enables to isolate the hydrodynamical effects and expose the richness of
the MHD formulation.

Finally, the model \texttt{NoPert} is a 2D representation of the 1D problem
assuming a homogeneous accretion stream and a static chromosphere.
It is the reference case in order to present the basic features of the shock
evolution.
We have chosen the value of $20\mathrm{G}$.

\section{Analytical approach}
  \label{sec:analytical}

\subsection{Description of the equivalent 1D problem}
  \label{sec:1Devolution}

The typical evolution of 1D accretion shocks, in the context of YSOs
\citep{Sac08,Sac10,Kol08}, can be summarized by discussing model
\texttt{NoPert}.
\begin{figure}
  \resizebox{\hsize}{!}{\includegraphics{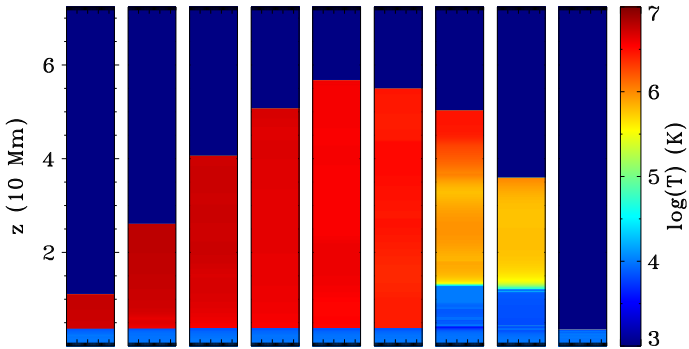}}
  \caption{Snapshots of the logarithmic temperature distribution during one
    oscillation of the reverse shock for the model \texttt{NoPert}.
    The dark blue color corresponds to the accretion stream, the light blue to
    the chromosphere and the red to the hot post-shock region which gradually
    cools down.
    The panels are separated by $108\,\mathrm{s}$, showing the formation and
    recollapse of the post-shock region due to the cooling instabilities induced
    by the optically thin radiation cooling.}
  \label{fig:ref_evol}
\end{figure}
\begin{figure}
  \resizebox{\hsize}{!}{\includegraphics{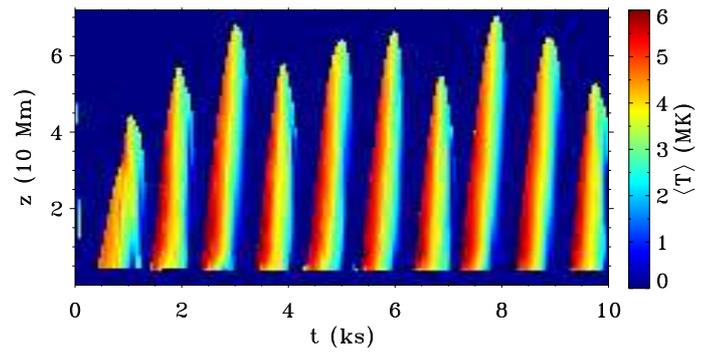}}
  \caption{The vertical structure of the emission-measure weighted temperature
    as a function of time.
    Cooling instabilities induce a quasi-periodic structure in the expected
    radiation with a period of roughly a thousand seconds.}
  \label{fig:ref_t-T}
\end{figure}
Figure~\ref{fig:ref_evol} shows the formation and collapse of the post-shock
region.
The impact of the accretion stream compresses the gas, heating it to
temperatures of a few $\mathrm{MK}$.
The post-shock region steadily increases in height as the reverse shock travels
upwards along the funnel.
However, the optically thin radiation losses of the hot slab are significant,
effectively cooling down the plasma.
Within the characteristic cooling time, the temperature drops by almost 3 orders
of magnitude and thus the pressure can no longer support the infalling gas.
In turn, the post-shock region collapses, repositing material onto the
chromosphere.
Subsequently, a new hot slab starts to build up and the process is repeated.
Since the X-ray emission depends on the density, temperature and volume of the
hot plasma, the emitted radiation of the accretion shock will show a
quasi-periodic temporal structure.

In fact, this can be seen in Fig.~\ref{fig:ref_t-T}, which displays the
quantity:
\begin{equation}
  \left<T(t,z)\right>
    = \frac{\int\rho^2T\,\mathrm{d}x}{\int\rho^2\,\mathrm{d}x}\,.
\end{equation}
This emission-measure weighted temperature is calculated at each $t$ and $z$ and
provides an estimate of the time dependence of the expected radiation
\citep{Orl10}.
Even though the integration in the $x$ direction is not important for this
$x$-symmetric simulation, it is particularly useful in the general 2D cases.
Evidently, the corresponding observable flux of model \texttt{NoPert} is
intermittent with an average periodicity of approximately a thousand seconds.
The small variation to the maximum height that the reverse shock reaches at each
oscillation, i.e. between $5$ and $7\,\mathrm{Mm}$, is attributed to the MHD
waves that propagate in the post-shock region and are reflected at the bottom
boundary and the shock surfaces.
These perturbations, that are independent of $x$, create density peaks in the
hot slab which in turn affect the time at which the cooling instabilities are
triggered.
For instance, the rightmost panels of Fig.~\ref{fig:ref_evol} show that the
vertical profile of the temperature during the collapse phase is variable.
Moreover, the cooling function adopted in our simulations is not smooth, as it
can be seen from Fig.~\ref{fig:coolfunc}, and hence its features in the presence
of waves may also contribute to producing nonidentical collapse phases.
Nevertheless, these are second order effects in the simulations and are
neglected for the rest of this study.

\subsection{Discussion on the qualitative effects of perturbations}
  \label{sec:qualitative}

In this context, the question that arises is to what extent perturbations can
disrupt this oscillatory stucture, and especially whether its periodic behavior
is suppressed.

The key quantity that determines the evolution of the system is the
characteristic cooling time, $\tau_\mathrm{cool}$.
The 1D approximation assumes that a unique value of $\tau_\mathrm{cool}$
describes the whole post-shock region, a fact that results in the quasi-periodic
global behavior described above.
However, if adjacent vertical flux tubes have either a small phase difference in
their oscillations or a slightly different $\tau_\mathrm{cool}$ value, then a
significant horizontal gradient in the pressure of the post-shock region would
develop due to the asychronous collapse.

In the case of a weak magnetic field, or equivalently in the high
plasma-$\beta$ regime, this will create large velocity components along the $x$
direction that subsequently will lead to a chaotic plasma motion.
The density inhomogeneities of the post-shock region will enhance the energy
losses, shortening also the cooling time \citep{Sut03}.
On the other hand, the presence of a relatively strong vertical magnetic field
will prevent horizontal interaction.
For instance, an out-of-phase collapse of neighboring flux tubes will still
retain their independent evolution\footnote{In the latter argument, we have
neglected communication by MHD waves because they do not affect
$\tau_\mathrm{cool}$ considerably.}.
Following \citet{Sac10} and \citet{Orl10}, who have briefly described this
mechanism, we refer to these magnetically confined vertical structures as
fibrils.
Hence, in the low plasma-$\beta$ case, these fibrils can be approximated as
isolated quasi-periodic emitters and the overall X-ray radiation is simply the
sum of their individual contribution.
Therefore, an out-of-phase oscillation would smooth out any global periodic
signature in the observed flux, even though each fibril would follow the 1D
quasi-periodic evolution described in sect.~\ref{sec:1Devolution}.

\subsection{Quantitative study of the fibril structure}

In order to quantify the problem we first derive some simplified expressions for
the fibril structure.

\subsubsection{The cooling time scale of fibrils}

A lower limit of the Mach number of the accreting gas is found by assuming a
relatively small infalling speed, $v_\mathrm{acc} = 100\,\mathrm{km\,s}^{-1}$,
for a gas of a temperature, $T_\mathrm{acc} = 1000\,\mathrm{K}$:
\begin{equation}
  M_\mathrm{acc} = \frac{v_\mathrm{acc}}{c_\mathrm{acc}} \simeq
    \frac{v_\mathrm{acc}}{\sqrt{5k_\mathrm{B}T_\mathrm{acc}/3\mu m_\mathrm{H}}}
    > 15\,,
\end{equation}
where $c_\mathrm{acc}$ is the sound speed in the accretion stream and
$k_\mathrm{B}$ the Boltzmann constant.
Given this high value, the shock can be generally assumed strong and hence its
postshock temperature is proportional to the velocity squared:
\begin{equation}
  T_\mathrm{sh} \propto v_\mathrm{acc}^2\,.
\end{equation}
Moreover, the optically thin cooling time, $\tau_\mathrm{cool}$, can be
estimated from Eq.~(\ref{eq:pr_eq}), since the plasma pressure changes by
$\sim$$P_\mathrm{sh}$ in that interval:
\begin{equation}
  \frac{P_\mathrm{sh}}{\tau_\mathrm{cool}} \propto \rho_\mathrm{acc}^2\Lambda
    \propto \rho_\mathrm{acc}^2T_\mathrm{sh}^{-1/2}\,.
\end{equation}
Here we have used the expression $\Lambda \propto T^{-1/2}$ as a rough
approximation in the temperature range $10^5$--$10^7\,\mathrm{K}$, see
Fig.~\ref{fig:coolfunc}.
Combining the above relations and given the fact that the kinetic energy of the
accretion stream is converted to thermal pressure upon impact,
$P_\mathrm{sh} \simeq \rho_\mathrm{acc} v_\mathrm{acc}^2$, the cooling time is
found to have the following dependence \citep[e.g.][]{Sac10}:
\begin{equation}
  \tau_\mathrm{cool} \propto \frac{v_\mathrm{acc}^3}{\rho_\mathrm{acc}}\,.
  \label{eq:tcool}
\end{equation}

\subsubsection{The length scales of the fibrils}
  \label{sec:fibwidth}

Let $h$ be the maximum height that the post-shock region attains and $d$ the
width of the fibrils.
The quantity $h$ can be estimated by multiplying the time that the reverse shock
propagates, $\tau_\mathrm{cool}$, with its velocity, which is proportional to
$v_\mathrm{acc}$ \citep[e.g.][]{Sac10}:
\begin{equation}
  h \propto v_\mathrm{acc}\tau_\mathrm{cool}
    \propto \frac{v_\mathrm{acc}^4}{\rho_\mathrm{acc}}\,.
\end{equation}
The width of the fibrils is relevant only in the low plasma-$\beta$ regime,
because otherwise the magnetic flux tubes cannot constrain the plasma dynamics,
and the infalling gas is effectively mixed after the impact.
In the former case, we can estimate $d$ from the horizontal component of
Eq.~(\ref{eq:v_eq}) by assuming a small deformation of the flux tube and
decomposing the magnetic field as $\vec B = b_x\hat x + (B_0 + b_z)\hat z$, with
$b_x, b_z \ll B_0$:
\begin{equation}
  \frac{\delta P}{\delta x}
    \simeq \frac{B_0}{4\pi}\left(\frac{\delta b_x}{\delta z}
    - \frac{\delta b_z}{\delta x}\right)\,.
\end{equation}
The first term of the right hand side represents the magnetic tension due to the
curvature of the magnetic fieldlines, $\delta b_x/\delta z\sim b_x/(h/2)$, and
the second term represents the gradient of the magnetic pressure.
We neglect the latter, since we do not expect a net difference in $b_z$ between
the inside and outside of the fibril.
Therefore,
\begin{equation}
  \frac{P_\mathrm{sh}}{d}
    \sim \frac{B_0}{4\pi}\frac{b_x}{h/2}
    \sim \frac{B_0^2}{8\pi}\frac{4d}{h^2}\,,
\end{equation}
where in the last relation we used the property of the tangent, i.e.
$b_x/B_0\simeq (d/2)/(h/2)$.
Finally, we obtain:
\begin{equation}
  d \propto \sqrt{\beta}\,h\,.
  \label{eq:fibsize}
\end{equation}
This expression relates the width of the fibril with its maximum height as well
as the plasma-$\beta$.

\subsubsection{Hydrodynamical limit, no fibrils}

In the high plasma-$\beta$ regime, the time scale for the homogenization of the
post-shock pressure is on the order of
$r_\mathrm{acc}/c_\mathrm{sh}\simeq500\,\mathrm{s}$.
Here we have assumed a fraction of the solar radius for the accretion stream
cross section, $r_\mathrm{acc}$$\sim$$10^5\,\mathrm{km}$, and a post-shock sound
speed of $c_\mathrm{sh}$$\sim$$200\,\mathrm{km\,s}^{-1}$.
This time scale is comparable to $\tau_\mathrm{cool}$ which suggests that any
potential phase differences between adjacent vertical elements of the hot slab
are likely to smooth out.
As a result, there are no fibrils forming and the accretion shock is expected to
maintain its periodic behavior in the high plasma-$\beta$ regime.

\subsection{The effects of perturbations}

In the context of the simplified configuration of Fig.~\ref{fig:sketch}, the
disruption of the global shock periodicity comes down to the search of a
potential mechanism that can introduce a net and adequate phase difference in
the oscillations of adjacent flux tubes, or equivalently a gradient in their
$\tau_\mathrm{cool}$ value.
In general, perturbations are characterized by three quantities, a) amplitude,
b) length scale and c) time scale.
Since any small phase difference between fibrils will not increase further in
the absence of perturbations, their amplitude cannot be arbitrarily small.
However, a continuous perturbation may have cumulative results, and thus even
small variabilities might be adequate, as long as they do not cancel each
other's effects in time.
Moreover, their spatial extent should be larger than the fibril width, $d$,
otherwise the contained post-shock region can be homogenized.
Finally, the temporal scale of the perturbation should be comparable to the
cooling time, $\tau_\mathrm{cool}$.

Below, we focus on the case of a strong magnetic field and we estimate the
effects of general types of perturbations (Fig.~\ref{fig:perturbation}) to the
temporal evolution of the shock structure.

\subsubsection{Density fluctuations in the accretion stream}

A radial density profile across the accretion column will inevitably destroy the
global periodicity due to the different resulting cooling times of each vertical
element (Bonito et al. in preparation).
However, here we do not account for such initial conditions, but instead we
study the perturbed uniform profile.
By considering small variations to the infalling gas density, since the velocity
is not expected to fluctuate considerably, Eq.~(\ref{eq:tcool}) gives:
\begin{equation}
  \tau_\mathrm{cool}
    \propto \frac{1}{\rho_\mathrm{acc} + \delta\rho_\mathrm{acc}}
    \simeq \frac{1}{\rho_\mathrm{acc}}
    \left[1 - \frac{\delta\rho_\mathrm{acc}}{\rho_\mathrm{acc}}
    + \left(\frac{\delta\rho_\mathrm{acc}}{\rho_\mathrm{acc}}\right)^2\right]\,.
\end{equation}
If the density perturbations are random, one has:
$\rho_\mathrm{acc}\pm\delta\rho_\mathrm{acc}$.
To first order, their effects cancel each other and $\tau_\mathrm{cool}$ remains
unchanged.
However, by taking second order terms into account, a net change does appear in
the cooling time.
Nevertheless, the average value of $\tau_\mathrm{cool}$ is affected globally,
i.e. at all $x$, and hence the system will always be in phase.
Consequently, we should rather focus on lasting phase shifts that can be
introduced between adjacent fibrils, namely a local variation of
$\delta\rho_\mathrm{acc}$.
Rewriting Eq.~(\ref{eq:tcool}) we obtain:
\begin{equation}
  \frac{\delta\tau_\mathrm{cool}}{\tau_\mathrm{cool}} \propto
    -\frac{\delta\rho_\mathrm{acc}}{\rho_\mathrm{acc}}\,.
\end{equation}
Therefore, a clump with a significant density constrast can considerably affect
the local cooling time, within a range comparable to its size.
Note that the presence of several clumps, apart from bringing the fibrils out of
phase faster, could also reduce the required amplitude of the perturbations due
to their cumulative effect.

\subsubsection{Perturbations in the chromosphere}

A variation of $\delta P_\mathrm{chr}$ and $\delta \rho_\mathrm{chr}$ may
relocate locally and temporarily the depth down to which the accretion shock can
penetrate into the chromosphere, as well as influence the post-shock cooling
efficiency.
Since pressure fluctuations
($\delta P_\mathrm{chr}$$\sim$$5\times10^3\,\mathrm{erg\,cm}^{-3}$) are larger
than the ram pressure of the accretion shock
($\rho_\mathrm{acc}v_\mathrm{acc}^2$$\sim$$300\,\mathrm{erg\,cm}^{-3}$), the
activity of the chromosphere is capable of influencing its evolution.
In fact, notice that even if we consider the solar energy flux, which is one
order of magnitude lower, the value of $\delta P_\mathrm{chr}$ would still be on
the same order as $\rho_\mathrm{acc}v_\mathrm{acc}^2$.
Moreover, this kind of variability is on the order of minutes with the typical
lengths involved being a few thousand $\mathrm{km}$
\citep[e.g.][]{Boh84,Jud06,Bel09,Bec09,Bec13}.
Therefore, it can act on the level of fibrils, rather than on the whole shock
structure, and also within the frequency window of the expected oscillations.
In other words, this type of perturbation has the appropriate amplitude, spatial
and temporal scales to potentially disrupt the synchronization and hence affect
the global quasi-periodic signature.

\subsubsection{Shape of the impacting surface}

Finally, even in the case of a homogeneous accretion stream and a static
chromosphere, the starting time for each fibril oscillation is determined by the
shape of the surface of first impact.
For instance, if the center of the funneled gas cross-section hits the star
earlier that its outer parts by an interval comparable to, or larger than,
$\tau_\mathrm{cool}$ then the formation and collapse of the fibrils will be out
of phase.

\section{Numerical study}
  \label{sec:results}

The previous rough estimates indicate that introducing perturbations in the
accretion shock system can potentially suppress the predicted periodic emission.
In this section, we explore the actual effects of several types of realistic
perturbations by performing numerical simulations and studying the
time-dependent shock structure.
We illustrate the involved processes in the dynamics, showing also the role of
the magnetic field.

\subsection{Random density perturbations in the stream}

\begin{figure}
  \resizebox{\hsize}{!}{\includegraphics{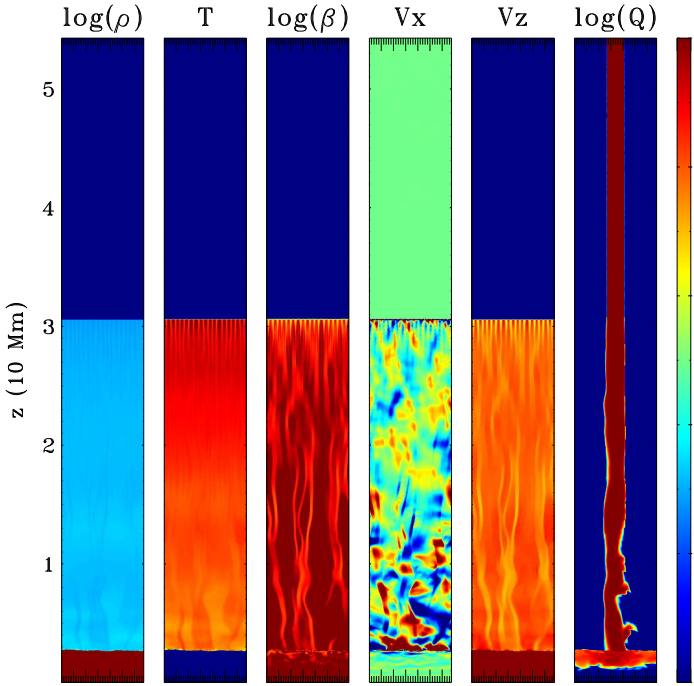}}
  \caption{Density, temperature, plasma-$\beta$, and velocity distributions at
    $t = 10\,\mathrm{ks}$ for the weak magnetic field model \texttt{Rnd20}.
    The colors blue/red are used for low/high values, and in particular
    $\rho\in[10^{-13},\,10^{-11}]\,\mathrm{g\,cm}^{-3}$,
    $T\in[0,\,4]\,\mathrm{MK}$, $\beta\in[10^{-2},\,10^2]$,
    $v_x\in[-5,\,5]\,\mathrm{km\,s}^{-1}$,
    $v_z\in[-500,\,0]\,\mathrm{km\,s}^{-1}$, and $Q\in[0,\,1]$.
    The $1\%$ density perturbations that are applied to this model are too weak
    to be visible.
    Due to the saturation of the density color, the exponential profile of the
    chromosphere cannot be seen.}
  \label{fig:rnd20_var}
\end{figure}
Figure~\ref{fig:rnd20_var} shows a snapshot of the spatial distribution of the
physical quantities for the high plasma-$\beta$ model \texttt{Rnd20}.
Three regions can be identified from top to bottom: the tenuous and cold
accretion stream, the denser post-shock region of a few million Kelvin, and the
relatively colder isothermal chromosphere which is described by an exponential
density profile.
As expected, the post-shock region resembles closely the unperturbed case
(\texttt{NoPert}) because the magnetic field is not strong enough to prevent
pressure fluctuations from being smoothed out.
This produces a chaotic motion of a few $\mathrm{km\,s}^{-1}$ as it can be seen
in the panel of $v_x$ whose extrema correspond to $\pm0.01v_\mathrm{acc}$.
The vertical velocity component of the shocked material has a similar signature
but this is not apparent in the corresponding plot due to the different color
range used ($\pm v_\mathrm{acc}$).
Furthermore, the accreted material is reposited all over the surface of the
chromosphere, as indicated by the tracer, $Q$.

\begin{figure}
  \resizebox{\hsize}{!}{\includegraphics{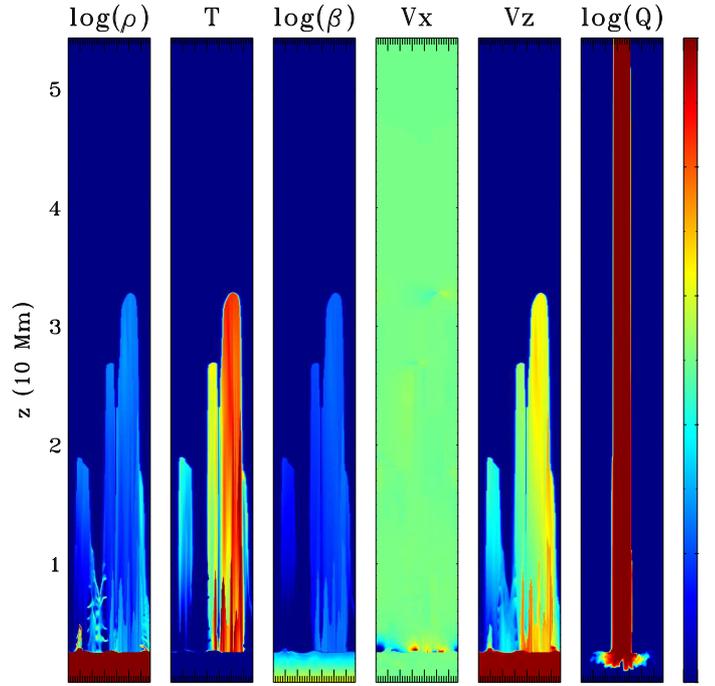}}
  \caption{Density, temperature, plasma-$\beta$, velocities, and tracer
    distibutions at $t = 10\,\mathrm{ks}$ for the strong magnetic field case
    \texttt{Rnd500}.
    The range of each plotted quantity is:
    $\rho\in[10^{-13},\,10^{-11}]\,\mathrm{g\,cm}^{-3}$,
    $T\in[0,\,2]\,\mathrm{MK}$, $\beta\in[10^{-2},\,10^2]$,
    $v_x\in[-5,\,5]\,\mathrm{km\,s}^{-1}$,
    $v_z\in[-500,\,0]\,\mathrm{km\,s}^{-1}$, and $Q\in[0,\,1]$.}
  \label{fig:rnd500_var}
\end{figure}
Figure~\ref{fig:rnd500_var} displays the same physical variables as
Fig.~\ref{fig:rnd20_var} but for the low plasma-$\beta$ model \texttt{Rnd500}.
Notice the fibril structure of the shocked material which is the result of the
magnetic confinement of the plasma inside vertical flux tubes.
The temporal evolution of each fibril is independent of its neighbours; some of
them are forming, e.g. regions on the right part of each panel, whereas others
are collapsing, e.g. regions on the left.
This can be seen by looking at the vertical distribution of the temperature of a
given fibril.
A forming fibril has a rather uniform temperature along $z$.
On the contrary, a collapsing fibril may have $T$ varying by orders of
magnitude along its height.
The horizontal speed component is zero almost everywhere, showing that plasma
mixing cannot take place easily in the low plasma-$\beta$ regime.
For the same reason, the tracer shows that the material accretes strictly
vertically and any further mixing can take place only deeper in the
chromosphere where the thermal pressure is stronger than the magnetic one.
Finally, note that despite the small perturbation amplitude, a phase shift does
develop among fibrils.

\begin{figure}
  \resizebox{\hsize}{!}{\includegraphics{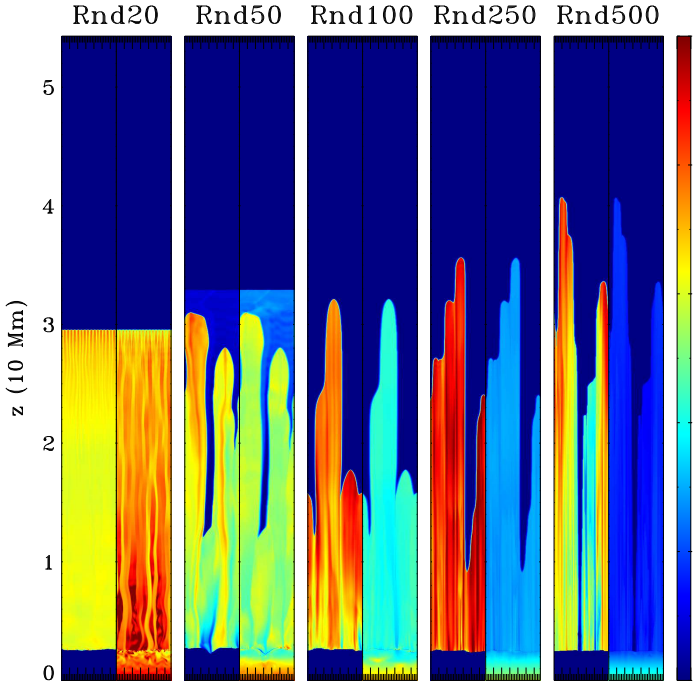}}
  \caption{Distributions of the temperature (on the left of each panel) and the
    plasma-$\beta$ (on the right of each panel) for the models \texttt{NoPert},
    \texttt{Rnd20}, \texttt{Rnd50}, \texttt{Rnd100}, \texttt{Rnd250}, and
    \texttt{Rnd500} (left to right).
    The plotted ranges are: $T\in[0,\,5]\,\mathrm{MK}$ and
    $\beta\in[10^{-3},\,10^3]$.
    The snapshots do not correspond to the same time, but each one was chosen
    such that the fibril structure becomes more apparent.
    Note also that the aspect ratio is not the same as the rest of the figures:
    these plots are compressed vertically by a factor of $1.5$.}
  \label{fig:fibril-beta}
\end{figure}
Figure~\ref{fig:fibril-beta} illustrates the dependence of the fibril size on
the strength of the magnetic field.
Although quantitative analysis is required to verify the approximate relation of
Eq.~(\ref{eq:fibsize}), namely $d\propto1/B_z$, it seems to be valid to zeroth
order (see Sect.~\ref{sec:fibwidth}).
Note that the size of the fibrils does not depend on the perturbation length
scale, which for these models is the cell width $\Delta x$, but only on the
plasma-$\beta$ value.
This is also confirmed by all other simulations of Table~\ref{tab:models}.
In addition, the fibrils seem to have a finer structure at the base close to the
chromosphere and a thicker and rounded shape at their head.

Furthermore, by assuming an accretion stream radius of $5\cdot10^4\,\mathrm{km}$
\citep{Orl10}, the computational box considered here represents approximately
one hundredth of its diameter.
This provides an estimate on the total number of fibrils that can form and also
shows the overall complexity of the post-shock region even in the case of
uniform magnetic fields.
On the other hand, we have verified that our simulations capture correctly the
width of the fibrils since it remains the same when increasing the resolution.
Nevertheless, lower plasma-$\beta$ values would demand smaller cell sizes due to
the anticipated thinner fibrils.
The above two points imply that simulating the entire stream together with the
highly structured post-shock region is a rather computationally expensive task.

\begin{figure}
  \resizebox{\hsize}{!}{\includegraphics{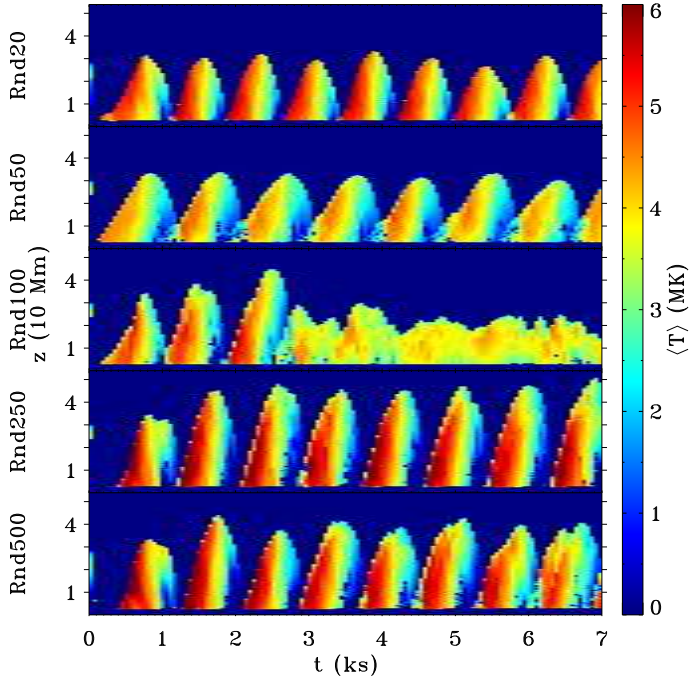}}
  \caption{Emission-measure weighted temperature versus $z$ and $t$ for the
    models that consider $1\%$ random density perturbations in the stream, i.e.
    \texttt{Rnd20}, \texttt{Rnd50}, \texttt{Rnd100}, \texttt{Rnd250}, and
    \texttt{Rnd500} (top to bottom).}
  \label{fig:strrnd-per}
\end{figure}
In order to quantify the effects that the $1\%$ random density fluctuations have
on the periodicity of the system, we plot in Fig.~\ref{fig:strrnd-per} the
emission-measure weighted temperature, $\left<T(z,t)\right>$, for this class of
models.
Overall, and despite the formation of fibrils in the low plasma-$\beta$ cases, a
strong periodic signature can be observed in the cases \texttt{Rnd20},
\texttt{Rnd50}, \texttt{Rnd250}, and \texttt{Rnd500}.
This is because the applied perturbation is small and it also has a zero net
effect along $x$.
In other words, fluctuations at each horizontal position cancel in time, making
each flux tube, on average, identical to the rest.
Recall that bringing the fibrils out of phase is not an unstable process and
therefore their oscillations remain generally synchronized with only a small
phase difference attributed to the small density fluctuations.
Here, a stronger and less symmetric perturbation is required to produce an
adequate phase shift among adjacent fibrils.

Nevertheless, model \texttt{Rnd100} seems to lose its periodic signature in the
long term.
Its plasma-$\beta$ at the post-shock region is $\beta \sim 1$ which suggests
that the dynamics are neither dominated by purely hydro processes nor by the
magnetic field alone, as it is the case for the other models.
Instead, there is a complex interplay between the two that gives rise to 2D
mechanisms which in turn disrupt the global periodic behavior even for such
relatively small perturbations.
Finally, by comparing Figs.~\ref{fig:ref_t-T} and \ref{fig:strrnd-per} we
observe that the maximum height attained by the perturbed cases is lower than
that of the unperturbed, and also that their oscillation period is slightly
shorter.
For the high plasma-$\beta$ cases this is due to the inhomogeneities produced in
the hot slab, where the locally denser regions are susceptible to stronger
energy losses.
Effectively, this reduces the average $\tau_\mathrm{cool}$ as it is also
observed and discussed in \citet{Sut03}.
On the other hand, even though the low plasma-$\beta$ cases do reach a higher
height than the weak $\vec B$ models, the small deformation of the flux tubes by
the confined hot plasma invalidates the $x$-symmetry and, as a result, the
fibrils do not attain the height of \texttt{NoPert}.

\subsection{Isolated clumps or fully clumped stream}

\begin{figure}
  \resizebox{\hsize}{!}{\includegraphics{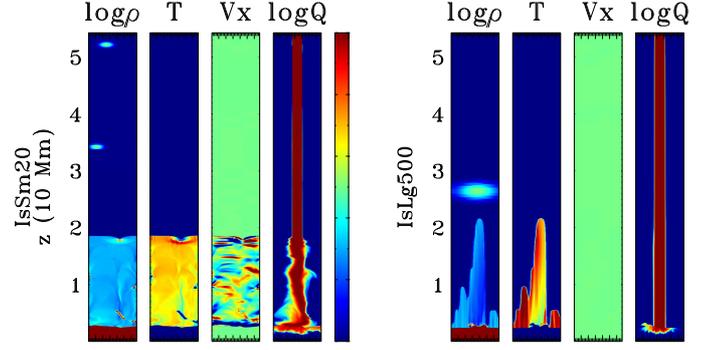}}
  \caption{Density, temperature, horizontal velocity, and the fluid tracer for
    the isolated clump models \texttt{IsSm20} (left panels) and \texttt{IsLg500}
    (right panels).
    Blue/red color corresponds to low/high values, i.e.
    $\rho\in[10^{-13},\,10^{-11}]\,\mathrm{g\,cm}^{-3}$,
    $T\in[0,\,5]\,\mathrm{MK}$, $v_x\in[-50,\,50]\,\mathrm{km\,s}^{-1}$, and
    $Q\in[0,\,1]$.
    The clumps, visible as elliptical spots in density distributions, are
    circular but appear elongated due to the compression of the vertical
    direction for visualization purposes.}
  \label{fig:strclis}
\end{figure}
Figure~\ref{fig:strclis} displays the physical quantities of two models that
incorporate isolated clumps into the stream, \texttt{IsSm20} and
\texttt{IsLg500}.
On the left, two small clumps are visible in the infalling gas and a third one
that has just hit the reverse shock.
The post-shock region is hotter where the clump interacts with the reverse
shock.
There, the basic physics is similar to that for the interaction of a supernova
remnant shock with a cloud of the ISM, where the major factor is the density
contrast of the clump with respect to the surrounding medium (e.g.
\citealt{Kle94}; see also \citealt{Orl08} for a MHD study including the effects
of anisotropic thermal conduction).
The shock-clump interaction generates transmitted (into the clump) and reflected
(into the post-shock material of the slab) shocks, whose temperature depends on
the density contrast of the clump.
The darkest portion of the shock front in the second panel of
Fig.~\ref{fig:strclis} is the shocked clump material and the red structure that
surrounds it is the bow shock.
The latter propagates into the shocked stream with a higher temperature than
that of its environment.
A strong velocity field appears locally, with the value of $v_x$ reaching almost
half of the accretion speed, $v_\mathrm{acc}$.
In fact, the hot slab shows overall much higher velocity values of chaotic
motion than the model \texttt{Rnd20}; here the velocity is plotted in the range
$\pm0.1v_\mathrm{acc}$ with larger values saturating to dark blue or dark red,
respectively.
On the other hand, the strong magnetic field of model \texttt{IsLg500} (on the
right) does not allow pressure homogenization and hence there is no plasma
mixing.
The formation of fibrils is again evident in the system, but this time their
oscillations are effectively brought out of phase.
This is due to the large density contrast of the clumps which significantly
modify the local value of $\tau_\mathrm{cool}$.

\begin{figure}
  \resizebox{\hsize}{!}{\includegraphics{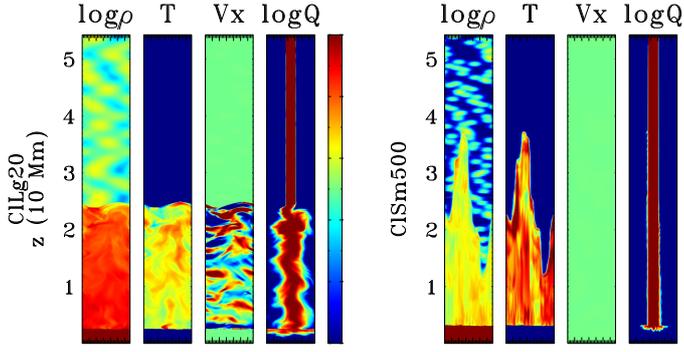}}
  \caption{Density, temperature, horizontal velocity, and the fluid tracer for
    models \texttt{ClLg20} (left panels) and \texttt{ClSm500} (right panels),
    which assume a fully clumped accretion stream.
    The plotted value ranges are:
    $\rho\in[10^{-14},\,10^{-12}]\,\mathrm{g\,cm}^{-3}$,
    $T\in[0,\,5]\,\mathrm{MK}$, $v_x\in[-50,\,50]\,\mathrm{km\,s}^{-1}$, and
    $Q\in[0,\,1]$.}
  \label{fig:strcldm}
\end{figure}
The general features observed in the isolated clump models are also seen in the
cases of fully clumped streams.
Figure~\ref{fig:strcldm} displays the spatial distribution of the physical
quantities for \texttt{ClLg20} and \texttt{ClSm500}.
In these cases too, the dynamics of the post-shock region depends on the
strength of the magnetic field and not on the type of perturbation.
Even though the average density of the infalling gas, and hence the cooling
time, is different than before, the temporal evolution is generally similar; the
high plasma-$\beta$ models show chaotic motion and mixing whereas the low
plasma-$\beta$ cases a fibril structure.

\begin{figure}
  \resizebox{\hsize}{!}{\includegraphics{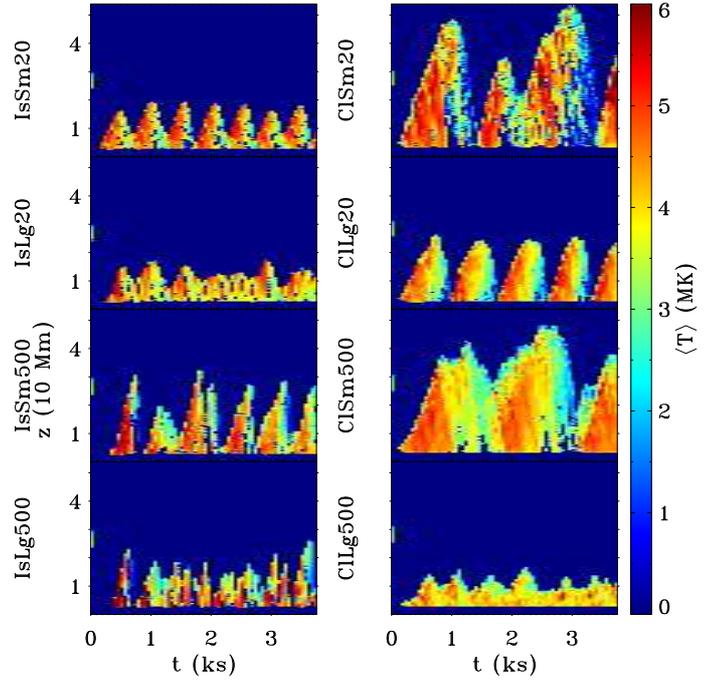}}
  \caption{Emission-measure weighted temperature for the models that assume
    isolated clumps (left panels) and the a fully clumped stream (right
    panels).}
  \label{fig:strcldm-t-T}
\end{figure}
The periodic signature of all the models that consider clumps can be derived
from Fig.~\ref{fig:strcldm-t-T}.
The presence of dense clumps increases the local cooling efficiency of the
post-shock region and as a result the emission may vary strongly with $z$.
The difference in the frequencies (when there is one) and the maximum height
attained are attributed to the value of the average density of the accreting
material.
In particular, the high plasma-$\beta$ models \texttt{IsSm20}, \texttt{ClSm20},
and \texttt{ClLg20} maintain a quasi-periodic oscillatory character.
On the other hand, the strong magnetic field models have their fibrils
oscillating fully out of phase.
As a result, the periodic behavior is heavily disrupted in most cases and
suppressed in the model \texttt{IsLg500}.
We note that the limited width and dimensionality of the box is contributing to
these noisy profiles.
If we had explored the whole accretion stream cross section or performed 3D
simulations, the sum of individual out-of-phase fibril contributions would have
provided a much smoother profile.

\subsection{Perturbations in the chromosphere}

\begin{figure}
  \resizebox{\hsize}{!}{\includegraphics{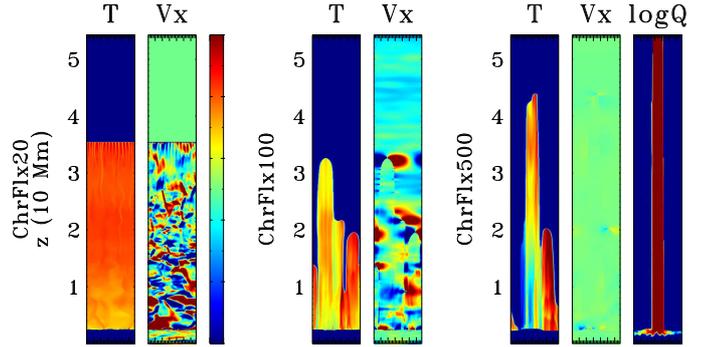}}
  \caption{Spatial distribution of the temperature and of the horizontal
    velocity component for the models that consider an energy flux in the
    chromosphere.
    Plotted ranges: $T\in[0,\,5]\,\mathrm{MK}$,
    $v_x\in[-5,\,5]\,\mathrm{km\,s}^{-1}$, and $Q\in[0,\,1]$.}
  \label{fig:chromo}
\end{figure}
Models \texttt{ChrFlx\#} show the same magnetic-field-dependent features of the
accretion shock structure that were discussed earlier, despite the intrinsically
different perturbation type.
The sinusoidal spatial profile of the chromospheric pressure fluctuations
generates local vertical velocities in all three cases and also a horizontal
motion when the magnetic field is weak enough.
As it can be seen in Fig.~\ref{fig:chromo}, the upwards travelling sonic waves
disrupt the lower edge of the post-shock region leading to a similar
$\vec B$-dependent morphology of the shocked material as the rest of the models
in Table~\ref{tab:models}.
The $v_x$ component is plotted in the range $\pm0.01v_\mathrm{acc}$ but we note
that the values of the chaotic motion can be by an order of magnitude larger
close to the chromosphere.
Furthermore, notice that \texttt{ChrFlx100} displays a non-zero value of $v_x$
in the stream.
This effect is created from MHD waves that travel along the accretion flow and
force the material to follow the wavy structure of the magnetic field lines.
Therefore, in the low plasma-$\beta$ regime, any type of chromospheric
fluctuations can also perturb the stream apart from the accretion shock.
This effect is present in all models, even though weaker.

\begin{figure}
  \resizebox{\hsize}{!}{\includegraphics{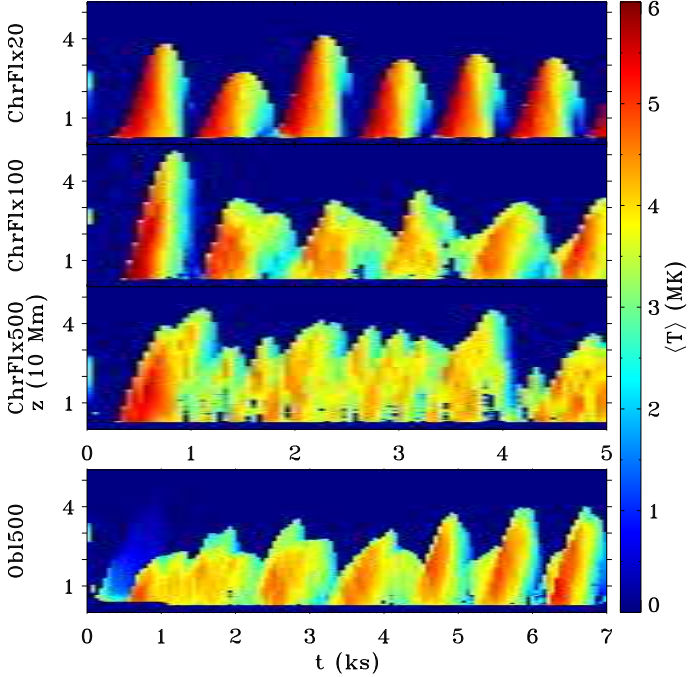}}
  \caption{Emission-measure weighted temperature for the models that consider
    fluctuations in the chromosphere (top three) and for the case of an oblique
    surface of impact (bottom).}
  \label{fig:chromo-t-T}
\end{figure}
The top three panels of Fig.~\ref{fig:chromo-t-T} show the emission-measure
weighted temperature versus time for the three models considering perturbations
of chromospheric origin.
Once again we confirm that the periodicity is present when the magnetic field is
weak.
In the same context, the independent fibril oscillations of the low
plasma-$\beta$ case hide any well-defined temporal patterns.

\subsection{Oblique impacting surface}

Finally, the bottom panel of Fig.~\ref{fig:chromo-t-T} presents model
\texttt{Obl500}.
By definition, this case is prepared such that among flux tubes there is a phase
shift on the starting time of the fibril oscillations.
However, once the whole stream has entered the box, no further perturbations are
introduced.
The plot suggests that although the fibrils are initially out of phase, they
gradually get synchronized.
This implies that a coordinated oscilation of the fibrils is a stable state and
that the weak interaction between the flux tubes favors the synchronization of
the system.

\section{Summary - conclusions}
  \label{sec:conclusions}

In the literature, theoretical studies and 1D numerical simulations predict an
oscillatory behavior in the emission of accretion shocks.
However, observational data appear not to confirm such an expectation.
In this paper, we have explored whether the presence of perturbations could
disrupt or even suppress this periodic signature.
We have performed 2D MHD simulations of the interior of the accretion flow to
investigate the effects of several realistic perturbations, such as a clumped
stream or fluctuations in the chromosphere.
We have explored a wide range of magnetic field values in order to understand
its role in the dynamics.
Our conclusions are summarized as follows:
\begin{itemize}
\item
The structure of the post-shock region is found to depend strongly only on the
plasma-$\beta$ value and only very weakly on the nature of the perturbation.
In particular, a relatively strong magnetic field confines the plasma in flux
tubes and leads to the formation of fibrils.
Their size depends on the value of $\vec B$ and their evolution follows roughly
a quasi-periodic oscillation that is almost independent of their neighbours.
In the high plasma-$\beta$ regime, the post-shock region shows chaotic motion
and plasma mixing with the value of local velocity determined by the applied
perturbation.
\item
In the cases where the magnetic field is weak, the emission maintains a strong
periodic signature.
This is because the pressure fluctuations that appear in the post-shock region,
due to the applied perturbations, are smoothed out and the entire hot slab forms
and recollapses as a whole.
However, both the oscillation frequency and the height reached by the reverse
shock are lower than those of the unperturbed case due to the more efficient
cooling, a byproduct of chaotic motion and the inhomogeneities it produces.
\item
For strong magnetic fields, the global periodicy in the emission can be
suppressed if the fibrils oscillate out of phase.
This requires a perturbation that can effectively modify the cooling time and/or
the trigger time among adjacent flux tubes.
We point out that since observations of protostars suggest high magnetic field
values, the low plasma-$\beta$ models we have investigated in this paper are
expected to be the most relevant for real YSOs.
\item
In general, for a perturbation to have an effect on the system, it has to have a
time scale comparable to the cooling time, a length scale comparable to the
fibril width, and an adequate amplitude.
All three criteria seem to be fulfilled by the typical perturbations expected to
be present in the accretion shock system.
\end{itemize}

In order to compare these theoretical accretion models with observations,
adequate radiative transfer post-processing treatment is required to determine
spectral signatures.
As part of further investigations, we will make use of the 3D IRIS code
\citep{Ibg13}, to predict the temporal, spatial, angular, and spectral
distributions of the radiation that emerges from such 2D/3D accretion shock
simulations.
It will be also interesting to address the possibility that the accretion flows
are fragmented, as recently suggested from solar observations \citep{Rea13}.

\begin{acknowledgements}
  We would like to thank an anonymous referee for the helpful comments on the
  manuscript.
  The authors are grateful to A.~Ciardi for his help and suggestions during the
  preparation of this study.
  We would also like to thank F.~Thais for his support as well as A.~Mignone,
  M.~Flock, S.~Matt, G.~Herczeg, J.~Bouvier, S.~Cabrit, A.~Maggio, A.~Bonanos,
  T.~Katsiyannis, R.~Keppens, C.~Xia, and R.~Pinto for fruitful discussions.
  This work was supported by the ANR STARSHOCK project
  (ANR-08-BLAN-0263-07--2009/2013) and was granted access to the HPC resources
  of CINES under the allocation 2012-c2012046943 made by GENCI (Grand Equipement
  National de Calcul Intensif).
\end{acknowledgements}

\bibliographystyle{aa}
\bibliography{paper}

\end{document}